\documentclass[sn-mathphys]{sn-jnl}
\usepackage{url,hyperref,lineno,microtype,subcaption}
\usepackage[onehalfspacing]{setspace}
\jyear{2023}%

\theoremstyle{thmstyleone}%
\theoremstyle{thmstyletwo}%

\theoremstyle{thmstylethree}%

\unnumbered

\begin{document}
\newcommand{\vdag}{(v)^\dagger}
\newcommand\aastex{AAS\TeX}
\newcommand\latex{La\TeX}
\newcommand{\ct}{$^{13}$C}
\newcommand{\ctb}{$^{13}$C~}
\newcommand{\cd}{$^{12}$C}
\newcommand{\cdb}{$^{12}$C~}
\newcommand{\apj}{{\it Astrophys. J.}}
\newcommand{\nphysa}{{\it Nuclear Physics A.}}
\newcommand{\apjl}{{\it Astrophys. J. Lett.}}
\newcommand{\apjs}{{\it Astrophys. J Suppl..}}
\newcommand{\pasp}{{\it Public. Astron. Soc,. of the Pacific}}
\newcommand{\pasa}{{\it Public. Astron. Soc. of Australia}}
\newcommand{\pasj}{{\it Public. Astron. Soc. of Japan}}
\newcommand{\mnras}{{\it Mon. Not. Roy. Astron. Soc.}}
\newcommand{\araa}{{\it Ann. Rev. Astron. Astrophys.}}
\newcommand{\actaa}{{\it Acta Astronomica}}
\newcommand{\physrep}{{\it Phys. Rep.}}
\newcommand{\aap}{{\it Astron. \& Astrophys.}}
\newcommand{\memsai}{{\it Mem. Soc. Astron. It.}}
\newcommand{\prl}{{\it Phys. Rev. Lett.}}
\newcommand{\gca}{{\it Geofisica Cosmofisica Acta}}
\newcommand{\prc}{{\it Phys. Rev. C.}}
\newcommand{\ssr}{{\it Space Sci. Rev.}}
\newcommand{\nar}{{\it New Astron. Rev.}}
\newcommand{\grl}{{\it Geophys. Res. Lett.}}
\newcommand{\nf}{$^{14}$N}
\newcommand{\nfb}{$^{14}$N~}
\newcommand{\ctan}{$^{13}$C($\alpha$,n)$^{16}$O}
\newcommand{\ctanb}{$^{13}$C($\alpha$,n)$^{16}$O~}
\newcommand{\nean}{$^{22}$Ne($\alpha$,n)$^{25}$Mg}
\newcommand{\neanb}{$^{22}$Ne($\alpha$,n)$^{25}$Mg~}
\newcommand{\fe}{$^{56}$Fe}
\newcommand{\feb}{$^{56}$Fe~}
\newcommand{\fes}{$^{60}$Fe}
\newcommand{\fesb}{$^{60}$Fe~}
\newcommand{\al}{$^{26}$Al}
\newcommand{\alb}{$^{26}$Al~}
\newcommand{\oq}{\textquotedblleft}
\newcommand{\cq}{\textquotedblright}
\newcommand{\cqb}{\textquotedblright~}
\newcommand{\ms}{M$_{\odot}$}
\newcommand{\msb}{M$_{\odot}$~}
\title[The heavy $s$-process]{Production of n-rich nuclei in red giant stars}


\author*[1]{\fnm{Busso} \sur{Maurizio}}\email{maurizio.busso@pg.infn.it}

\author[2, 1]{\fnm{Palmerini} \sur{Sara}}\email{sara.palmerini@pg.infn.it}
\equalcont{These authors contributed equally to this work.}

\affil*[1]{\orgdiv{Section of Perugia}, \orgname{INFN}, \orgaddress{\street{Via A. Pascoli snc}, \city{Perugia}, \postcode{I-06123}, \country{Italy}}}

\affil*[2]{\orgdiv{Department of Physics and Geology}, \orgname{University of Perugia}, \orgaddress{\street{Via A. Pascoli snc}, \city{Perugia}, \postcode{I-06123},  \country{Italy}}}


\abstract{We outline a partial historical summary of the steps through which the nucleosynthesis
phenomena induced by {\it slow} neutron captures (the {\it s-process}) were clarified, a scientific
achievement in which Franz K\"appeler played a major role. We start by recalling the early phenomenological approach, which yielded a basic understanding of the subject even before models for the parent stellar  evolutionary stages were developed.  Through such a tool, rough limits for the neutron density and exposure were set, and the crucial fact was understood that more than one nucleosynthesis component is required to account for solar abundances of $s$-process nuclei up to the Pb-Bi region. We then summarize the gradual understanding of the stellar processes actually involved in the production of nuclei from Sr to Pb (the so-called {\it Main Component}, achieved in the last decade of the past century and occurring in red giants of low and intermediate mass, {\it M} $\lesssim$ 8 \ms) populating, in the {\it HR} diagram, the {\it Asymptotic Giant Branch} or {\it AGB} region.  We conclude by giving some details on more recent research concerning mixing mechanisms inducing the activation of the main neutron source, \ctan.}

\keywords{Nucleosynthesis: neutron captures, the $s$-process, stars: evolution of, stars: Asymptotic Giant Branch, solar abundances}



\maketitle

\section{Introduction}\label{sec1}

Several decades have passed since the first fundamental papers on stellar nucleosynthesis by Burbidge, Burbidge, Fowler and Hoyle \cite{b2fh} (hereafter $B^{2}FH$) and by Cameron \cite{cam57} appeared. 
They had been preceded by the discovery of the radioactive element Tc in old red giant stars by \cite{merrill}, by the compilation of meteoritic abundances by Suess and Urey \cite{su56}, as well as by crucial contributions on stellar nuclear reactions and an outline of the nuclear shell model \cite{green54, cory52}. These ingredients formed the basis for the general sketch of nucleosynthesis processes in stars outlined by $B^2FH$. According to their discussion, neutron-induced reactions are even now generally classified as
being {\it slow} ($s$-process) or {\it rapid} ($r$-process), depending on their average rates as compared 
to those of the typical competing $\beta$-decays of unstable insotopes involved
(for a recent outlook, see e.g. \cite{th23}). 

A series of works after $B^2FH$ concentrated in particular on the first of the above processes, characterized by a path following closely the valley of $\beta$ stability, hence by a possibility of being, broadly speaking, clarified even knowing neither the details about the nuclear structure (masses and decay rates) of nuclei far from stability, nor those of the astrophysical models for the producing stars. Clayton et al. \cite{clay1} and Seeger et al. \cite{sfc} were the first to lay down the mathematical tools of the so-called {\it phenomenological approach} to the $s$-process, also outlining relevant results that were bound to permit important steps forward in the understanding 
of neutron captures. 

The above authors recognized the crucial role played, for clarifying the synthesis processes of nuclei beyond iron, by the solar $\sigma_i N_i$ distribution (products of the cross sections for (n,$\gamma$) reactions on isotopes $-i$ times their solar abundances). Any theoretical formulation of these products could then be compared to the improved available set of data for the solar system composition \cite{urey64} and for the neutron-capture cross sections \cite{mg65}.  

From this basic work, a first understanding of the problems involved in slow neutron captures emerged, which will be briefly recalled in Section 2. Starting from the late sixties, stellar models began to afford quantitatively the advanced stages of stellar evolution for low mass stars ($LMS, M$/\msb $\lesssim 3 -4$) and for intermediate-mass stars ($IMS, 4 \lesssim M$/\msb $\lesssim 8-9$), the so-called {\it Asymptotic Giant Branch}, or $AGB$, stages. Here \cite{sh} discovered how He burning in a thin shell (bracketed by the H-rich envelope above and the C-O degenerate core below) is prone to repeated thermal instabilities (see also \cite{sf78, pri81, mow99}), inducing sudden flashes of energy release and the development of intermediate convective zones sweeping the whole inter-shell layers (they are often referred to as {\it thermal pulses}). Nucleosynthesis occurring in those unstable regions became an important subject for stellar physics in the Seventies, especially after \cite{ul73} showed the possible direct correlation between thermal He-shell instabilities hosting neutron captures (these last induced by the \neanb reaction)
and repeated neutron exposures. In this field, the group led by Icko Iben at Urbana (Il) soon took a leadership role \cite{bi80, ib82, ti77, kib78, it78}. These topics will be briefly recalled in Section 3. A crisis then followed, induced by the understanding that $s$-process rich $AGB$ stars are usually of  low mass ($M \lesssim 3 - 4$ \ms) and have therefore rather low temperatures even in the He-shell instabilities ($T \lesssim 3 \times 10^8 K$), so that  the \neanb reaction operates at a very low efficiency. The ensuing works by several groups (and especially by researchers operating in Torino), looking for the activation of the alternative neutron source \ctanb, will be outlined in Section 4. These studies, in particular, were performed in close collaboration with Franz K\"appeler and colleagues, as described in another paper of the present collection, authored by R. Gallino \cite{gal23}. Then, in Section 5, we shall outline more recent efforts, done in the last twenty years in various places (and in particular in our University, in Perugia), looking for physical models of mixing processes capable of overcoming the oversimplified parameterizations affecting the modelling of the neutron source \ct, now seen as dominating neutron captures in $AGB$ stars. Some encouraging results and preliminary conclusions will also be illustrated in Section 6.

\section{The phenomenological approach}\label{sec2}

As mentioned, the first outline of a mathematical formulation for the $s$-process was due to \cite{clay1} and \cite{sfc} and is now generally referred to as the {\it phenomenological approach} to neutron captures. The analysis by those authors started from the consideration that, in the simplest possible representation of neutron captures occurring on two adjacent stable nuclei of atomic mass $A$ and $A-1$, the abundance by number 
$N(A)$ of the nucleus $A$ varies in time as:

\begin{equation}
\frac{dN(A)}{dt} = N(A-1) n_n <\sigma(A-1)\cdot v> - N(A) n_n <\sigma(A)\cdot v> \label{eq1}
\end{equation}
\noindent
where $<\sigma(A) \cdot v >$ is the Maxwellian-averaged product between the neutron-capture cross section for the nucleus $A$ and the plasma velocity, while $n_n$ is the neutron density. One can then define the {\it neutron exposure} $\tau$, i.e. the time integrated neutron flux, to be:

\begin{equation}
\tau = \int_0^t{n_n v_T dt}
\end{equation}

\noindent
where $v_T$ indicates the thermal velocity. By subsequently introducing the $average$ cross section, $\hat{\sigma}(A)$:

\begin{equation}
\hat{\sigma}(A) = \frac{< \sigma \cdot v>}{v_T}
\end{equation}
\noindent
equation (\ref{eq1}) becomes:

\begin{equation}
\frac{dN(A)}{d\tau} = N(A-1) \hat{\sigma}(A-1) - N(A) \hat{\sigma}(A) \label{eq2}
\end{equation}
When the neutron flow is  in steady-state conditions (i.e. $dN/d \tau = 0$), equation (\ref{eq2}) yields:
\begin{equation}
\hat{\sigma}(A) N(A) = const.
\end{equation}
\noindent
which tells how, if slow neutron captures occur at equilibrium, then we expect that the resulting solar $\sigma N$ products of stable nuclei affected only by neutron captures remain constant. 

The above condition is actually met (roughly) experimentally, for isotopes that lay far from the regions
where the shell model foresees the closure of neutron shells, in which jumps in the nuclear-charge radius occur and very stable nuclear structures are found. This last situation characterizes nuclei that contain specific numbers of neutrons, namely $N = 2, 8, 20, 28, 50, 82, 126$ (called {\it magic numbers}). 

\begin{figure}[t!!]
\includegraphics[width=1.2\linewidth]{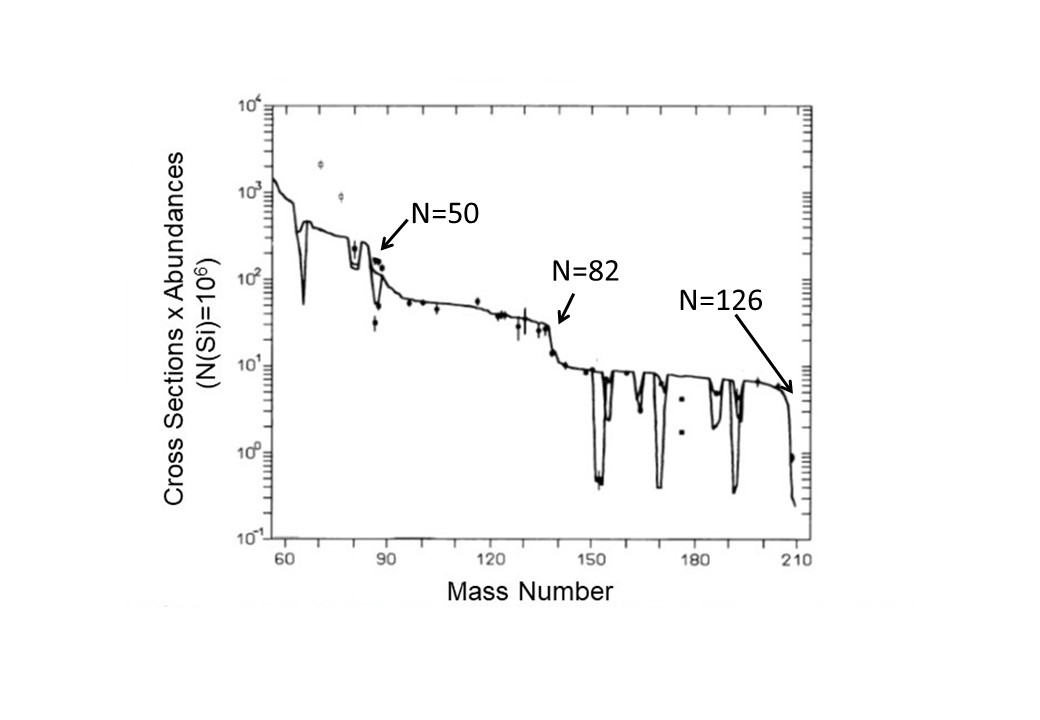}
\caption{The solar $\sigma N$ distribution, with the cross sections estimated at the reference temperature
of 30keV [dots], as available in 1990 \citep[adapted from][]{kae90}. The solid curve refers to the phenomenogical approximation described in the text, extended to include also nuclei depending on reaction branchings and using an {\it exponential distribution} of neutron exposures.}
\label{Fig1}
\end{figure}

These jumps are, for example, evident for $N =50, 82, 126$, in Figure \ref{Fig1}. They correspond to nuclei that reveal their high stability through relatively large abundances and very small cross sections (few millibarns). More generally, the same authors \cite{sfc} showed how the experimental solar products $\sigma \cdot N$ could be mimicked by distributions of neutron exposures decreasing for increasing values of $\tau$, like e.g in a power-law with a negative exponent or in a negative exponential form. In particular, they suggested to approximate the experimental data through the curve:

\begin{equation}
\rho(\tau) = \frac{GN^{56}_{\odot}}{\tau_0} exp(-\tau/\tau_0) \label{eq6}
\end{equation}

\noindent Here, a fraction $G$ of the solar $^{56}$Fe (the main nucleus on which $n$-captures occur) $N^{56}_{\odot}$ is subject, in the $s$-process site, to an exponential series of neutron irradiations and $\tau_0$ is a free parameter (called {\it mean neutron exposure}). Subsequently, Clayton \& Ward \cite{cw74} found that, adopting expositions like in equation (\ref{eq6}), the $\sigma \cdot N_s$ products ($N_s$ being the $s$-process fractional abundance of the various isotopes) could be expressed analytically by the formula: 
\begin{equation}
\sigma N_s = \frac{GN^{56}_{\odot}}{\tau_0} \prod_{i=56}^{A}\left( 1 + \frac{1}{\sigma_i \tau_0} \right)^{-1}
\end{equation}
It became also clear that, in the above conditions, more than one astrophysical mechanism would be required to overcome the \oq bottlenecks \cqb represented by magic nuclei and those authors suggested 
a superposition of three exponential {\it components}, namely a \oq weak\cqb one, for nuclei up to Sr (with a number of neutrons $N \lesssim 50$), a \oq main\cqb one, for nuclei above Sr and lighter than Pb (with $50 \lesssim N \lesssim 126$) and a \oq strong\cqb one, mainly requird for the magic $^{208}$Pb.

A treatment for branching points, where the chain of neutron captures meets $\beta$-unstable isotopes and a competition between capture and decay becomes possible, was then cosidered by \cite{wn78}, thus opening the possibility to estimate the environmental conditions (neutron density and temeprature) in the original stellar sites, from estimates of the {\it branching ratio} $f_{-}$:
\begin{equation}
f_{-} = \frac{\lambda^-}{(\lambda^- + \lambda_n)}
\end{equation}
\noindent
where $\lambda^{-} = 1/\tau_{\beta^{-}}$ is the rate of beta decays and $\lambda_n = n_n <\sigma\cdot v >$ is the rate of neutron captures.

From the above purely nuclear considerations, and remembering that many $\beta$-decay rates depend on the temperature $T$, four characterizing parameters ($G, \tau_0, n_n$ and $T$) could be inferred, thus providing tools useful to guide astrophysicists in their search for suitable stellar models. 

\section{Early nucleosynthesis models for $AGB$ stars}
Soon after the hypothesis of exponential distributions of exposures as drivers of $s$-processing was advanced by \cite{sfc}, stellar models were extended to the final evolutionary 
stages of $LMS$ and $IMS$, noting the \oq peculiar\cqb behavior of the $AGB$ stages powered by
nuclear burning of H and He, in a double-shell structure.

Here \cite{sh} underlined a peculiarity strongly affecting $s$-processing, called {\it thin shell instability}.
It occurs when, in  spherical  symmetry, a constant mass $\Delta m$ is concentrated at a constant radius $r$ in a thin shell of thicknes $l << r$ ($\Delta m \simeq 4 \pi \rho r^2 l$) and is in hydorstatic equilibrium, so that:
\begin{equation}
\frac{dP}{P} = -4\frac{dr}{r}
\end{equation}
One then has:
\begin{equation}
\frac{d\rho}{\rho} = -\frac{dl}{l} \simeq -\frac{dr}{l} = -\frac{r \cdot dr}{l \cdot r}
\end{equation}
Hence:
\begin{equation}
\frac{dP}{P} \simeq 4 \frac{l}{r }\frac{d\rho}{\rho}
\end{equation}
In the stellar plasma one also has:
\begin{equation}
\frac{dP}{P} = \alpha\frac{d \rho}{\rho} + \beta\frac{dT}{T}
\end{equation}
Hence:
\begin{equation}
\left(4\frac{l}{r}-\alpha\right)\frac{d\rho}{\rho} = \beta \frac{dT}{T}
\end{equation}
For thermal stability one would require 
\begin{equation}
4\frac{l}{r} > \alpha
\end{equation}
\noindent
which, for $l/r \rightarrow 0$, is no longer satisfied. This drives an unstable situation. Suppose indeed that the layer expands. If the pressure from hydrostatic equilibrium drops with radius faster than that due to the expansion, then the layer will continue to expand
and cool out of equilibrium. If not, then the temperature will rise, inducing a thermal instability and an ensuing sudden enhancement of the luminosity, because of a {\it thermonuclear runaway}, before relaxation occurs reestablishing the previous condition. See e.g. \cite{str23} for a discussion.

\begin{figure}[t!!]
\includegraphics[width=1.2\linewidth]{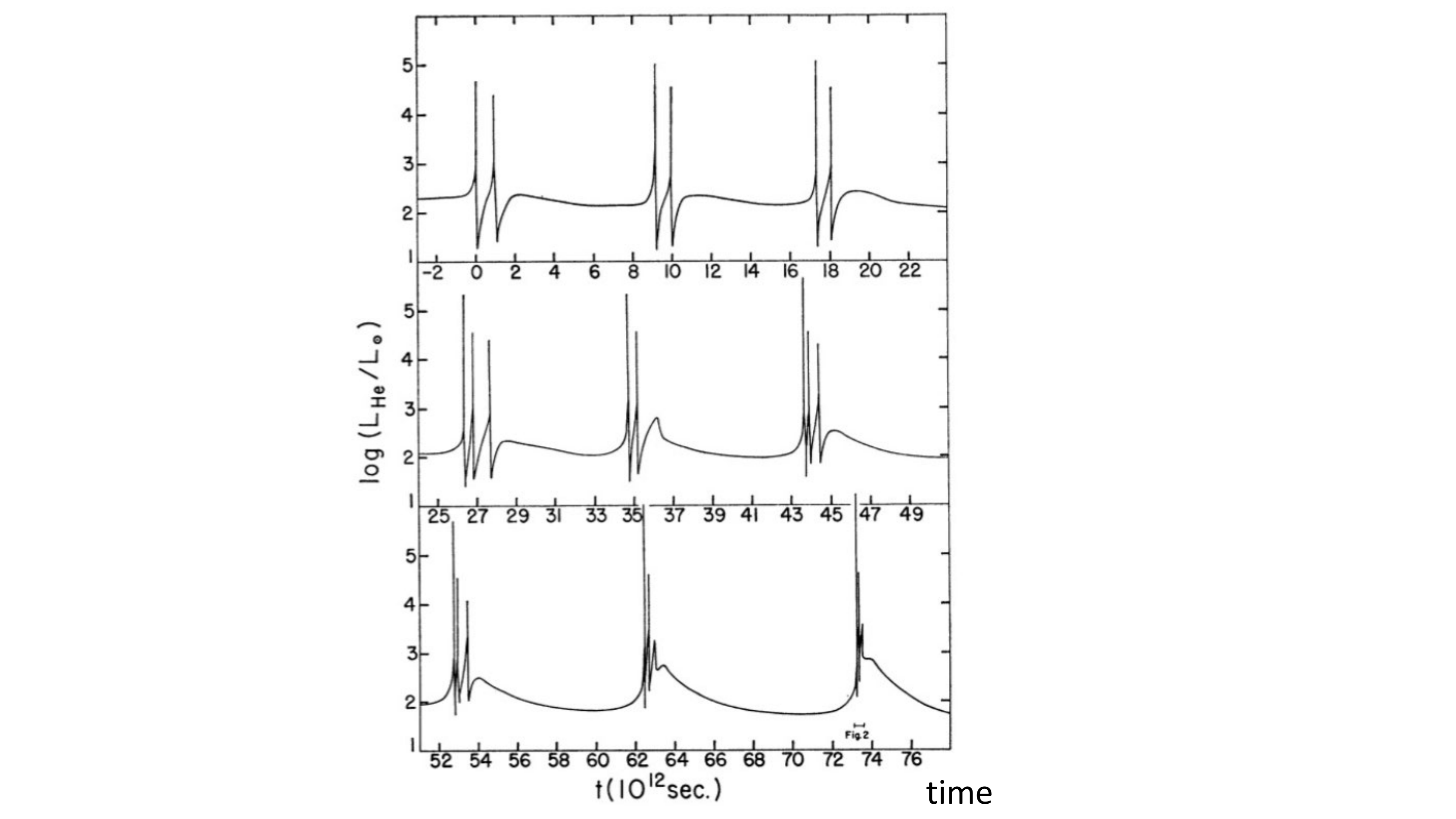}
\caption{The subsequent occurrence of instabilities in the 
He-shell luminosity, as a consequence of thermal flashes (pulses) experienced by He burning in a thin shell above a degenerate 
C-O core, in the work by \cite{sh} (adapted from the original paper).}
\label{Fig2}
\end{figure}

The repeated occurrence of the above unstable condition during the evolution of the star is shown in Figure \ref{Fig2}, adapted from the original paper by \cite{sh}. As a consequence of that peculiarity, $LMS$ and $IMS$ in their final stages of evolution (the so-called {Thermally-Pulsing-Asymptotic Giant Branch, or {\it TP-AGB} stages) are characterized by recurrent sudden luminosity flashes, accompanied by 
the development of Intermediate Convective Zones ($ICZ$), each followed by the penetration of the convective envelope below the upper border of the He-rich layers (see Figure \ref{Fig3}, adapted from  a famous review by Iben and Renzini \cite{ir83}).
\begin{figure}[t!!]
\includegraphics[width=1.2\linewidth]{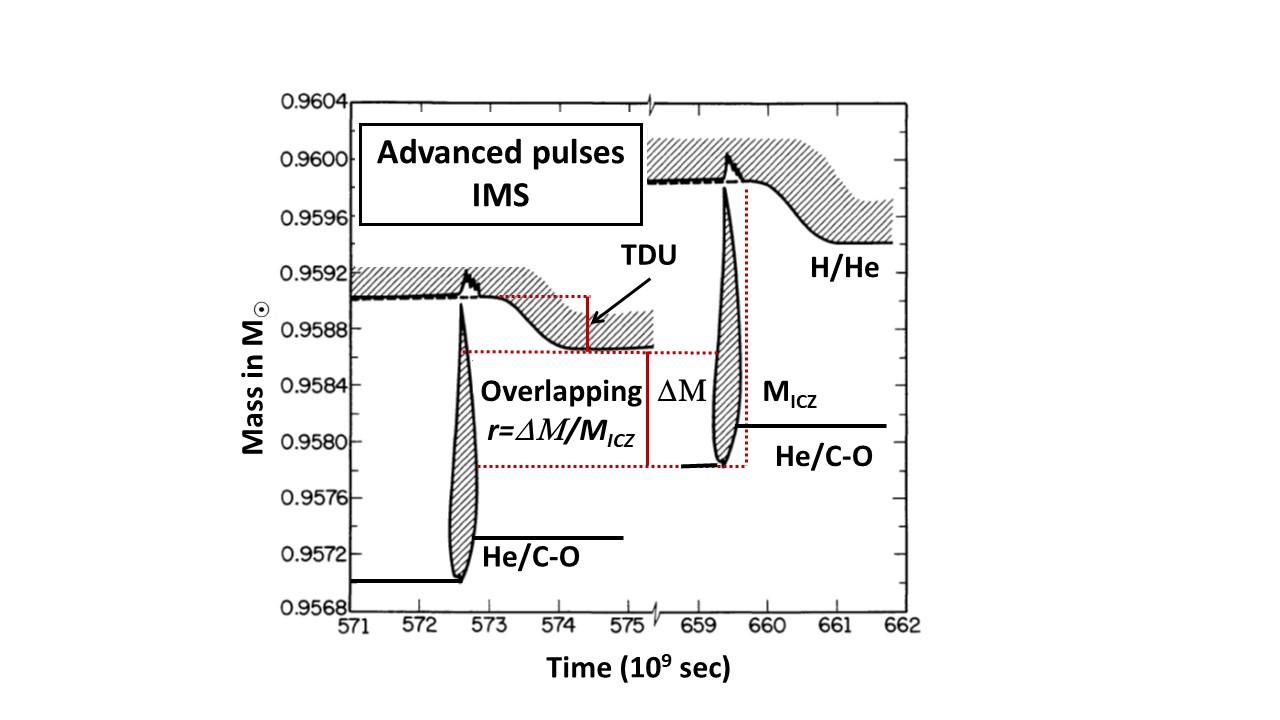}
\caption{A sketch of two succesive {\it thin shell} thermal instabilities of an $IMS$, developing $ICZs$, showng the overlapping between them  and the occurrence of the {\it Third Dredge-Up}. (See text foir details.)} \label{Fig3}
\end{figure}
As the figure shows, He-burning (which is normally inhibited by the core degeneracy) restarts abruptly (in a semi-explosive way) 
after long periods of H-burning activity that compresses and gradually heats the He-rich layers. Hence, He-re-ignition drives the mentioned instability, in which a peak in temperature occurs (starting from 1 $-$ 1.5 10$^8$ K and reaching up to $3 - 3.5$ $10^8$ K, mainly depending on the stellar mass). 

The large energy output thus generated (see also Figure \ref{Fig2}) induces the star to develop an $ICZ$ to increase the efficiency of the energy transport, so that the products of He-burning are spread throughout the intershell layers. These last then expand and cool, with the H-burning shell switched off,  and the convective envelope penetrates below the H-He discontinuity in what is called the {\it Third Dredge Up, or TDU}, carrying to the surface parts of the He-burning ashes (in particular carbon). The lower masses of the interested mass range, i.e. those from about 1.5 \msb to about 3$-$4 \ms, mix into the envelope enough $^{12}$C to make the C/O ratio become larger than unity (these are the so-called {\it Carbon Stars}, see the review by Straniero et al. \cite{str23} in this same volume).

The temperature achieved at the $ICZ$ base ($T ranging between 2.8 and 3.5$ $10^8$ K) is in general sufficient to drive further processes of $\alpha$ capture. In particular, from the abundant $^{14}$N left by CNO cycling in H-shell burning, the chain:

\begin{equation}
^{14}N(\alpha,\gamma)^{18}F(\beta^+\nu)^{18}O(\alpha,\gamma)^{22}Ne
\end{equation} 
can be driven. If the temperature is high enough ($T \gtrsim 3 \times 10^8 K$) this is followed by the reaction:
\begin{equation}
^{22}Ne(\alpha,\gamma)^{26}Mg
\end{equation} 
and by the competing:
\begin{equation}
^{22}Ne(\alpha,n)^{25}Mg \label{eq17}
\end{equation} 
with a higher probability for second process. 

Reaction (\ref{eq17}), in particular, makes intense neutron fluxes available, so that for almost two decades it became natural to assume that the $s$-process was originated in this way, thus explaining the original observations of Tc by Merrill \cite{merrill} in an $AGB$ star. 

The above conclusion seemed to receive a nice confirmation when Ulrich \cite{ul73} noted how the exponential distributions of exposures $\rho(\tau)$, suggested by \cite{sfc}, might be well accounted for during He-shell instabilities. This can be easily illustrated by considering the neutron exposition $\Delta \tau$ produced by the \neanb reaction in each  $ICZ$ of mass $M_{ICZ}$, overlapping with the subsequent one by a factor $r$ = $\Delta M_{ICZ}/M_{ICZ}$. Indeed, by neglecting the local effect of dredge-up and indicating with $\tau_0$ the ratio $\Delta \tau/(- \ln(r))$, it is simple to derive that:
\begin{equation}
\rho(\tau) \propto \frac{(1-r)}{\Delta \tau} r^{\tau/\Delta \tau} \propto \frac{1}{\tau_0}exp(-\tau/\tau_0) 
\end{equation}
which has the same form as in equation (\ref{eq6}).

The general properties of {\it TP-AGB} stars, crossing the phases where {\it thin shell instabilities} are met, were studiend by Paczy\'{n}ski in the early seventies \cite{pac1, pac2, pac3}, deriving analytical relations linking the main stellar parameters (like the interpulse period, the luminosity, the rate of advancement of the H-burning shell), usually expressing them as a function of the H-exhausted core mass, $M_H$. Similar relations are usually easily derived in stellar models, thanks to the convergence of evolutionary tracks to a narrow range of parameter values, ultimately due to the common property of $LMS$ and $IMS$ of witnessing only the evolution of a rather thin layer above  a degenerate pre-white-dwarf core of a mass below the Chandrasekhar limit.

\begin{figure}[t!!]
\includegraphics[width=1.2\linewidth]{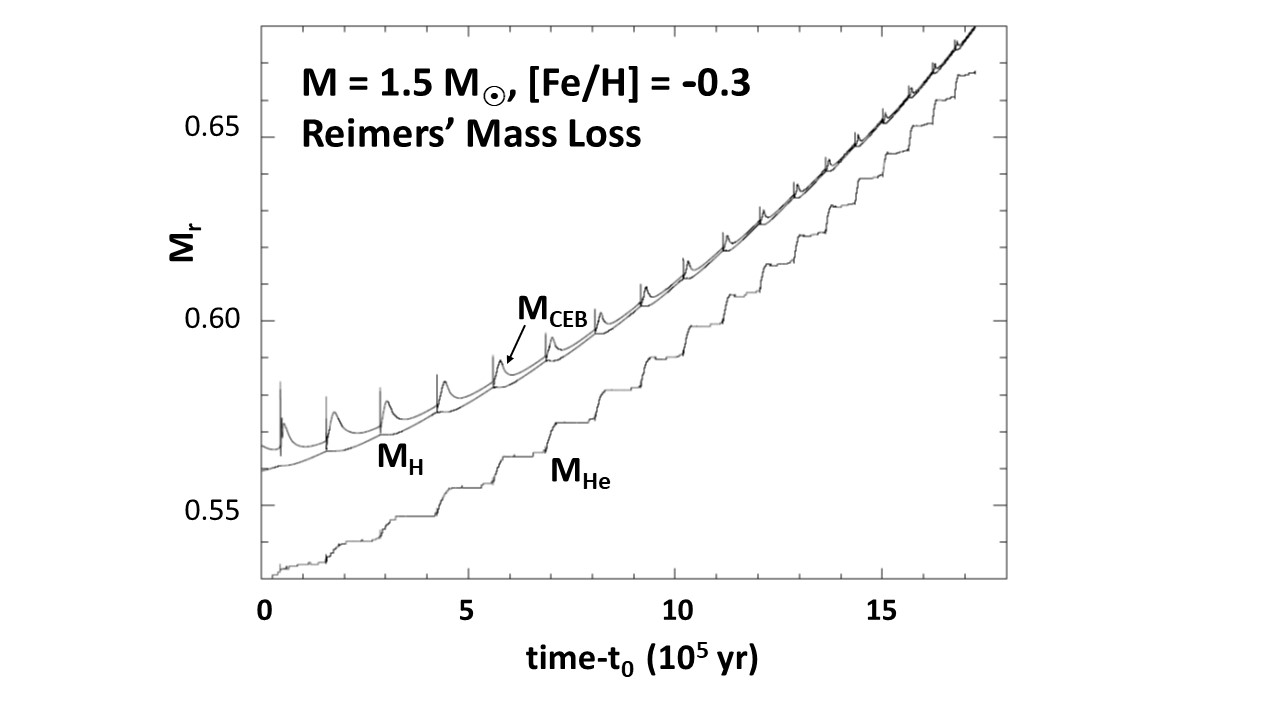}
\caption{The evolution as a function of time  of the position in mass of the Convective Envelope Bottom ($M_{CEB}$), of the interface between H-rich and He-rich layers (i.e. of the H-exhausted core, $M_{\rm H}$) and of the He-exhausted core ($M_{\rm He}$) in the model of a 1.5 \ms, half-solar metallicity star, obtained through the FRANEC evolutionary code. \label{Fig4}}

\end{figure}

The typical structure of the layers outside the CO core, for a low-mass $AGB$ star crossing these evolutionary stages to become a C-star \cite{str23}, is shown in Figure \ref{Fig4}. The figure was obtained with the software FRANEC (Frascati RAphson-Newton Evolutionary Code, see e.g. Chieffi \& Limongi \cite{cl} and Straniero et al. \cite{str, str97}). There, we recall that the notation [Fe/H] refers to:
$$
{\rm [Fe/H]} = \log \left[{\frac{N(\rm Fe)}{N(\rm H)}}\right]_{star}-\log\left[ {\frac{N(\rm Fe)}{N(\rm H)}}\right]_{\odot}
$$
The fact that $s$-process-rich C stars observationally tended to appear as being of low mass (despite the huge uncertainties then affecting this kind of measurements and in contrast with the needs of the \neanb source) was noticed by the same Icko Iben as early as in 1981 \cite{i81b} in a famous paper, after several years dedicated by his group to study more massive $AGBs$ \cite{ti77, it78,i81a}. As a conseqence, he and A. Renzini also suggested a possible way to initiate neutron captures from the alternative \ctanb source
\cite{ir82}, thus pioneering subsequent trends of the research.

Initially, the same accounting for $TDU$ episodes allowing freshly synthesized nuclei to be mixed into the envelope was a serious problem of stellar models, at least for low mass stars ($M \lesssim 3$ \ms). This was strictly connected to the choice of opacity tables. Negative results on $TDU$ were common, see e.g. \cite{bs88a, bs88b}, while positive outcomes were instead found by  \cite{i75}, \cite{w81} and later by \cite{john86, john89}. Here numerical problems (like model inadequacies
e.g. in the mesh rezoning), insufficient opacity tables, and the specific choices for the mixing algorithms at the envelope border all played crucial roles. 

The lengthy calculations required by $TDU$ with the computing facilities then available made the use of semi-analytical models for {\it TP-AGB} phases rather common, a method that
proved in general effective and provided in the years important results
\cite{paola96, wg98, paola22}.   

A growing evidence for the mixing of $s$-process elements to the surface was  in the mean time emerging, after the original Merrill's discovery. Data accumulated rapidly for all the classes of polluted {\it TP-AGB} stars, i.e. $MS, S, SC, C(N)$ giants, clarifying the gradual enrichment in heavy nuclei due to dredge-up along the {\it TP-AGB} evolutionary stage (see e.g. 
\cite{gust89, str23} for reviews widely separated in time).

Before the end of the Eighties, in general, it had become unquestionable that $AGB$ stars enriched in $s$-elements were of low mass, and hence could not have derived their enrichments from neutron captures induced
primarily by the \neanb source, due to the too low temperature at the base of $TP$s in those objects \cite{b+88}. The missing evidence for any variation in Mg isotopes at their surface confirmed that idea \cite{lambert91}.

Today, this conclusion has been certified beyond any uncertainty, thanks to the {\it Luminosity Function} of Galactic carbon stars, obtained trough the GAIA high-accuracy parallaxes \cite{a+22}. These data confirm their luminosities 
as being in line with theoretical predictions for $AGB$ stars below about 3 \msb ($M_{bol} \simeq - 5$).

\section{$s$-processing with the \ctanb reaction} \label{sec4}

In the same year of the negative conclusion drawn on \neanb as an origin for $s$-processing in AGB stars by \cite{b+88}, a short note by \cite{g+88} examined the possibilities of the altenative \ctanb source. 

Since the amount of \ctb left by H-shell burning is minimal in comparison to what is required for a neutron exposure like the one expected by the phenomenological approach (see section \ref{sec2}), these authors assumed the idea by \cite{ir82} that a further mixing of protons from the envelope could be induced below the formal convective border at each $TDU$, by variations in the carbon opacities at the low temperatures achieved in the expansion after a thermal pulse. The fact that repeated extra-mixing episodes could enhance the amount of \ctb in red giants had indeed been ascertained several years before by \cite{d+75}.

The mentioned analysis led to the conclusion that neutron production by \ctanb with sufficient amounts of \ctb available, was adequate to reproduce the solar distribution of $s$-process abundances. This however required to discover suitable extra-mixing mechanisms (parameterized for the moment) and to derive cross sections for all the relevant nuclei at the low temperature ($T \lesssim 1.5 \cdot 10^8$ K, $\simeq 12$k eV) of \ctb burning in thermal pulses, being the rough scaling with thermal velocities then available ($\sigma v_T =$ cost) rather inaccurate.

The ensuing need for revising many ideas on $s$-processing, as well as many experimental estimates, led naturally to a collaboration between the Torino and Karlsruhe groups, started with a paper by \cite{k+90} and continued till Franz K\"appeler's premature demise. Although several independent contributions came obviously from other groups in the subsequent decade (\cite{i91, bt91, b+92, bt97, mja98}), the mentioned collaborating effort stimulated many of the subsequent theoretical and experimental results, and produced some of the most widely cited reviews (\cite{b+99, k+11}).

In particular, from a thorough recomputation of the relevant stellar models, with the inclusion of updated physical parameters and opacity tables, \cite{str} clarified that any \ctb produced by newly mixed protons at $TDU$ would normally be consumed locally, at a temperature $T \lesssim 10^8$ K ($\simeq 8$ keV), in the radiative phases that preceed the next thermal pulse, except in rare conditions and at very low metallicities \cite{c+09}. The models born out of this revision produced important results, subsequenly applied for interpreting $s$-enriched AGB giants and their descendants \cite{b+95, b+01a, b+01b, a+01, a+02, a+03} as well as low metallicity carbon-enhanced metal-poor stars \cite{bi+10, bi+11, bi+12} and the renewed set of isotopic abundances discovered in presolar SiC grains of AGB origin \cite{liu1, liu2, liu3}. The main features of the mentioned models are summarized in \cite{g+98, b+99,
ar+99}.

The upgrade of the nuclear data for $s$-processing was in the mean time pursued, especially by the new maeasurements of the CERN facility $n_{-}TOF$ \cite{k00, bor+02a, bor+02b, ab+03} and summarized in a series of dedicated compilations \cite{bao00, k03, k1}.

\section{Mechanisms for extra-mixing} \label{sec5}

As we have seen, since the paper by \cite{ir82} it became clear that activating the neutron source \ctanb requires extensive mixing beyond the oversimplified treatment of convection usually adopted in canonical stellar models. 

Actually, in those years various types of non-convective mixing had been recognised to occur in Red Giant Stars ($RGB$) \cite{dv, hw03a, hw03}, especially to account for the evolution in the abundances of nuclei like $^3$He, $^7$Li, CNO isotopes and $^{26}$Mg (from the decay of $^{26}$Al).  
In most cases, this was a consequence of the observational evidence that, in the photosphere of $RGB$ stars, $^7$Li, C-O isotopes and other nuclei vary as a consequence
of increasing mixing of H-burning products \cite{g89, gb91, ch94, edl, p97}. 

The mechanisms explored for the above scopes were not included in standard stellar models, and required a separate, long and dedicated work for discovering the involved physics and its time scales 
\cite{c+12,ms17,dm,palm11a,palm11b}. 

Among the processes explored, we can 
mention atomic diffusion  \cite{mi86}, shears induced by rotation \cite{cv92,pier13}, {\it thermonhaline} diffusion related to local unbalance in the moleculare weight $\mu$, this last in most cases due to $^3$He burning  \cite{edl, cz07, cl10}. In general, in hydrostatic phases of stellar evolution, most of the above phenomena are very slow and require millions of years for producing effects. This fact excludes them as possibilities for producing the neutron
source \ct, as the duration of $TDU$ is just of a few centuries (normally from 1 to 4, depending on the stellar mass and pulse number).

\subsection{Extramixing associated to $TDU$}\label{sec5p1}

Broadly speaking, we call for time variations of the mass density $\rho$, through diffusion or 
circulation movements. In order to analyze what  is really needed at $TDU$, let us then rewrite the equation of state of a perfect 
gas as:

\begin{equation}
\rho = \left(\frac{m_H}{k_B}\right) \left(\frac{P}{T}\right) \mu
\end{equation}

where $k_B$ is the Boltzmann's constant, $m_H$ is the atomic mass unit, $\mu$ is the molecular weight and $P$ and $T$ indicate pressure and temperature,
respectively. In order to have a variation of $\rho$ with time, we can act either
on the molecular weight $\mu$ or on the $P/T$ ratio.

Most processes affecting $\mu$ are slow, except in explosive conditions \cite{dm}. Hence,
we must find something that can change the $P/T$ ratio at rates slower than convection, 
but much faster than for the other knowm extra-mixing mechanisms. So far, it appears that the most 
promising phenomena for doing that are those related to wave-like perturbations, 
like e.g Internal Gravity Waves ($IGW$) \cite{dt, bat19} or magnetic instabilities  (\cite{mestel}). Both types of phenomena are known to occur in red, evolved stars (\cite{bow+19, vle11, vle11b, vle12}).

In particular, since it is almost evident that complex magnetic fields must be generated in complex 
astrophysical plasmas, the Perugia group recently explored magnetic instabilities as possible causes of mixing.

The idea was advanced in \cite{b+07}, with an application to $AGB$ stars of early models for the Sun by E.N. Parker \cite{Park58, Park60}. This was then numerically
verified to be plausible by \cite{n+08}. The original paper presented semi-quantitative motivations 
to argue that, for magnetic fields of the order of the solar ones, magnetized bubbles would float at a 
speed fast enough to transport effectively matter across the envelope 
border during a $TDU$ episode, but a real confirmation of this fact had 
to wait for magneto-hydrodynamical ($MHD$) models.

These last started from the consideration that the force balance per unit mass, in a magnetized 
stellar plasma (see e.g. \cite{mestel}), can be written as: 

\begin{equation}
\frac{D {\vec v}}{D t} = 
\frac{\partial {\vec v}}{\partial t} + {{\vec v}\cdot {\vec \nabla}} {\vec v} = 
\vec{F}-\frac{1}{\rho}{\vec \nabla}p - {\vec \nabla}\Phi + \mu {\vec \nabla}^2 {\vec v}+\frac{1}{4 \pi \rho} {\vec B} \times ({\vec \nabla}\times{\vec B})
\end{equation}
(see e.g. \cite{nb14}). Here the Lorentz's term satisfies the known identity:
\begin{equation}
\frac{1}{4 \pi \rho} {\vec B} \times ({\vec \nabla}\times{\vec B}) = \frac{1}{4 \pi \rho} ({\vec B} \cdot {\vec \nabla}) {\vec B} - \frac{1}{\rho}{\vec \nabla}\left(\frac{B^2}{8 \pi}\right)
\end{equation}
that clarifies how, in equation (20), the presence of a magnetic field introduces two new terms, the gradient of an extra pressure ($B^2/8\pi$) and a tension, which acts in keeping the field confined into {\it flux tubes}. Due to the pressure gradient term, these tubes are pushed outward and forced to float. What had to be ascertained was actually the fact that the speed of the buoyant flux tubes was sufficiently fast to permit mixing within the short time duration of a $TDU$ episode.

\subsection{Magnetic buoyancy as applied to $AGB$ $s$-processing}\label{sec5p2}

There is today a large observational evidence that magnetic fields 
persist in low mass $AGB$ stars and in white dwarfs \cite{sk03,jor05,herpin,kem07,bl20}. 
Hence, in the induction equation:
\begin{equation}
\frac{\partial{\vec B}}{\partial t} = {\vec \nabla} \times (\vec{u} \times \vec{B}-\eta[\vec \nabla] \times \vec{B}) 
\end{equation}
(where $\eta$ is the Ohmic diffusivity coefficient) a velocity field $\vec{u}$ must be devised, 
having inductive properties suitable to substain $\vec{B}$ against dissipative forces. That
is the stellar {\it dynamo problem}, discussed in several works, e.g. in \cite{char}. 

For our purposes, let's remember the simple model by \cite{nb14}, based on a purely toroidal field. There, the geometrical simplification permits to derive exact $3D$ analytical solutions for the buoyant magnetic field. In them, the radial component of the buoyancy velocity can be written as:

\begin{equation}
v_r = \frac{dw}{dt} r(t)^{-(k+1)}
\end{equation}
while the toroidal componeent of the field ${\vec B}$ (the only original one) is:
\begin{equation}
B_{{\phi}} = \Phi({\xi}) r(t)^{k+1}
\end{equation}
where:
\begin{equation}
\xi(t) = w(t) + r(t)^{k+2}
\end{equation}

$\Phi$ and $\xi$ being arbitrary functions and $k$ being the exponent of the relation $\rho  \propto r^k$. The funtions must be specified on the basis on the particular problem afforded. For example, if:
\begin{equation}
w(t) = \Gamma t
\end{equation}
with: 
\begin{equation}
\Gamma = v_p r_p^{k+1}
\end{equation}
one has:
\begin{equation}
v_r = v_p \left(\frac{r_p}{r}\right)^{k+1}
\end{equation}

where $r_p$ and $v_p$ indicate radius and buoyant velocity at layer where buoyancy starts. The number $k$ can be derived from the physical
structure of He-rich layers immediately below convection at $TDU$. In the He-rich layers it turns out 
to be always $k\lesssim -4$. This implies a buoyancy velocity growing with radius more rapidly than a cubic parabola.

From the above solution, Trippella et al. \cite{tri} showed that, below the $TDU$ bottom sited at 
$r = r_e$, a penetration of protons would occur, with a mass distribution described by the relation:

\begin{equation}
\Delta M_p \simeq X_p\frac{4\pi\rho_E}{\alpha}{[r^2_e-\frac{2}{\alpha}r_e+\frac{2}{\alpha^2}]-
[r^2_p-\frac{2}{\alpha}r_p+\frac{2}{\alpha^2}] e^{-\alpha(r_e-r_p)}} 
\end{equation}

Here, $\rho_e$ is the average gas density at the convective border and $X_p$ is the fractional mass of protons in the envelope. In the formula, the $\alpha$ parameter can be adjusted  by limiting proton pollution to zones where the viscosity is small enough to permit the treatment by \cite{nb14} of an almost ideal MHD. This in turn depends on the stellar parameters of the model under study, with the polluted mass being normally around 0.004 $-$ 0.005 \ms.

\begin{figure}[t]

\centering
\includegraphics[width=0.7\textwidth]{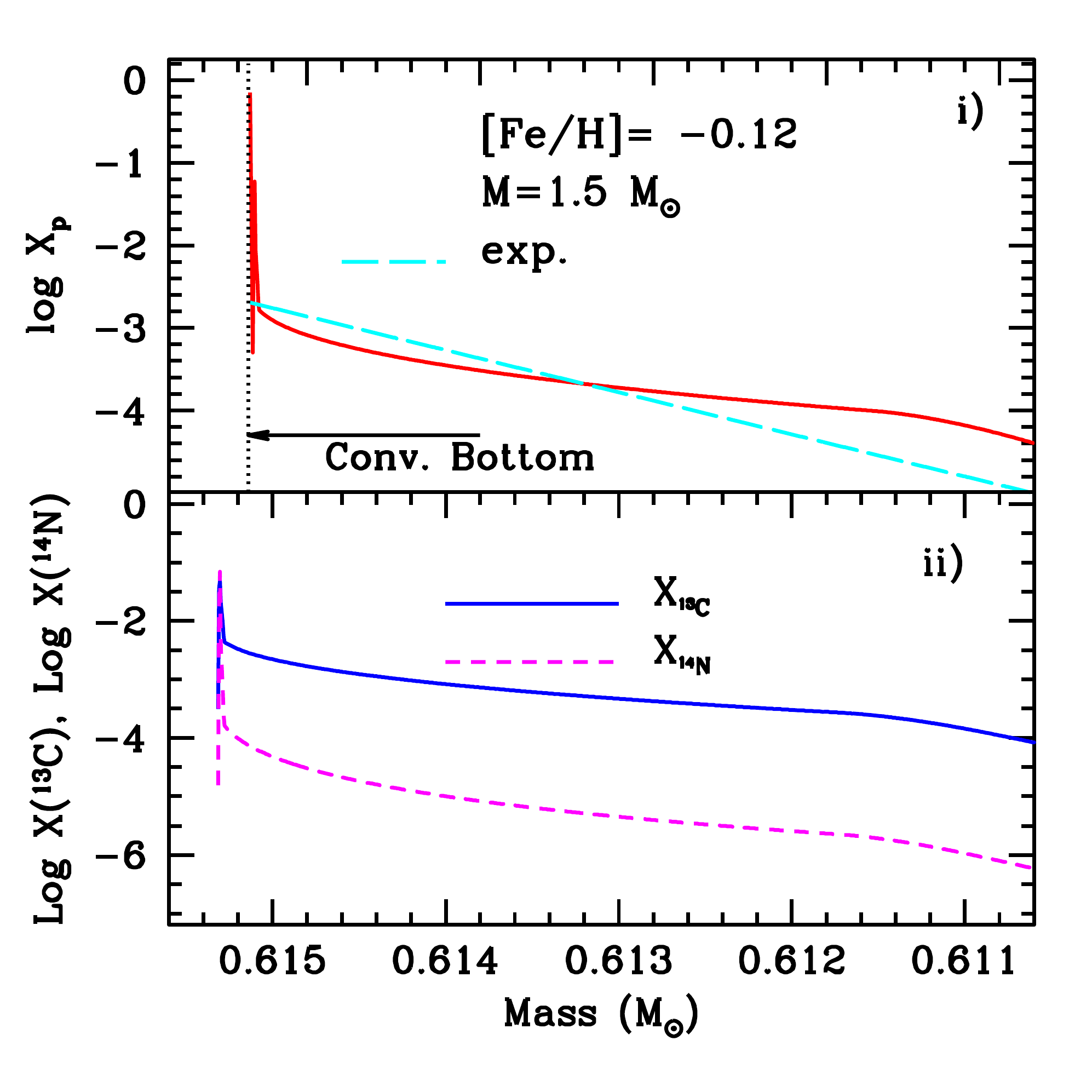}
\caption{Panel i): the reservoir ({\it pocket}) of protons as mixed from the envelope,
by magnetic buoyancy, computed according to \cite{nb14, tri}. Panel ii): the ensuing distribution of \ctb and $^{14}$N, later generated by H-burning. The Figure refers to the nineth $TDU$ episode of a 1.5 \msb $AGB$ star of about
solar metallicity. The dashed line in panel i) indicates the profile that would be obtained, for the same mixed mass, with a simpler explonental behavior in mass.}   
\label{figpock}      
\end{figure}

An example of the ensuing abundance of protons in the mixed region  is illustrated in Figure \ref{figpock} (i). At H reignition above these layers, the abundances of \ctb and of $^{14}$N are shown in Figure \ref{figpock} (ii) as a functon of the mass.

\section{Encouraging results and future work.} \label{sec6}

Nucleosynthesis models for Galactic stars, computed by assuming the presence, at each TDU episode,
of a \ctb reservoir as in equation (29), were presented 
recently by \cite{v+20, b+21}. When adopting nuclear inputs for neutron-capture cross sections and for 
weak interaction rates as discussed in \cite{p+21} and after averaging over standard choices for the 
Initial Mass Function and for the Star Formation Rate, it is possible, for such models, to reproduce quite well the 
solar abundance distribution of heavy elements and other common observational constrains from $s$-process enriched sites in the Galaxy \cite{b+21, b+22} (see e.g. Figures \ref{fig-6}-\ref{fig-8}).

\begin{figure}[t!!]
\centering
\includegraphics[width=0.75\textwidth]{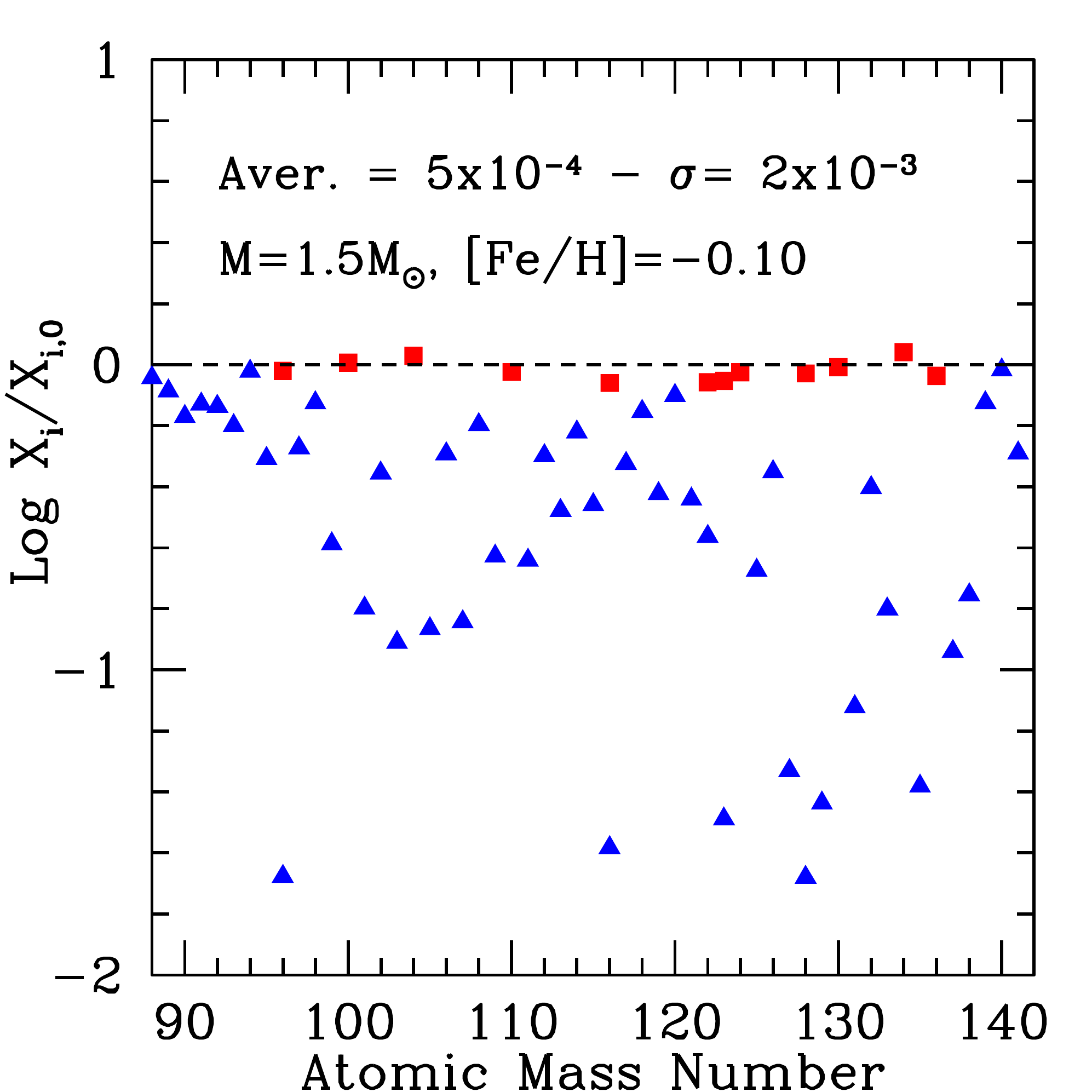}
\caption{Predictions of the solar fractional abundance distribution from the models of stellar $s$-process nucleosynthesis discussed in the text, for nuclei from $^{88}$Sr to $^{142}$Ce.}   
\label{fig-6}      
\end{figure}

\begin{figure}[ht]
\centering
\includegraphics[width=1.1\textwidth]{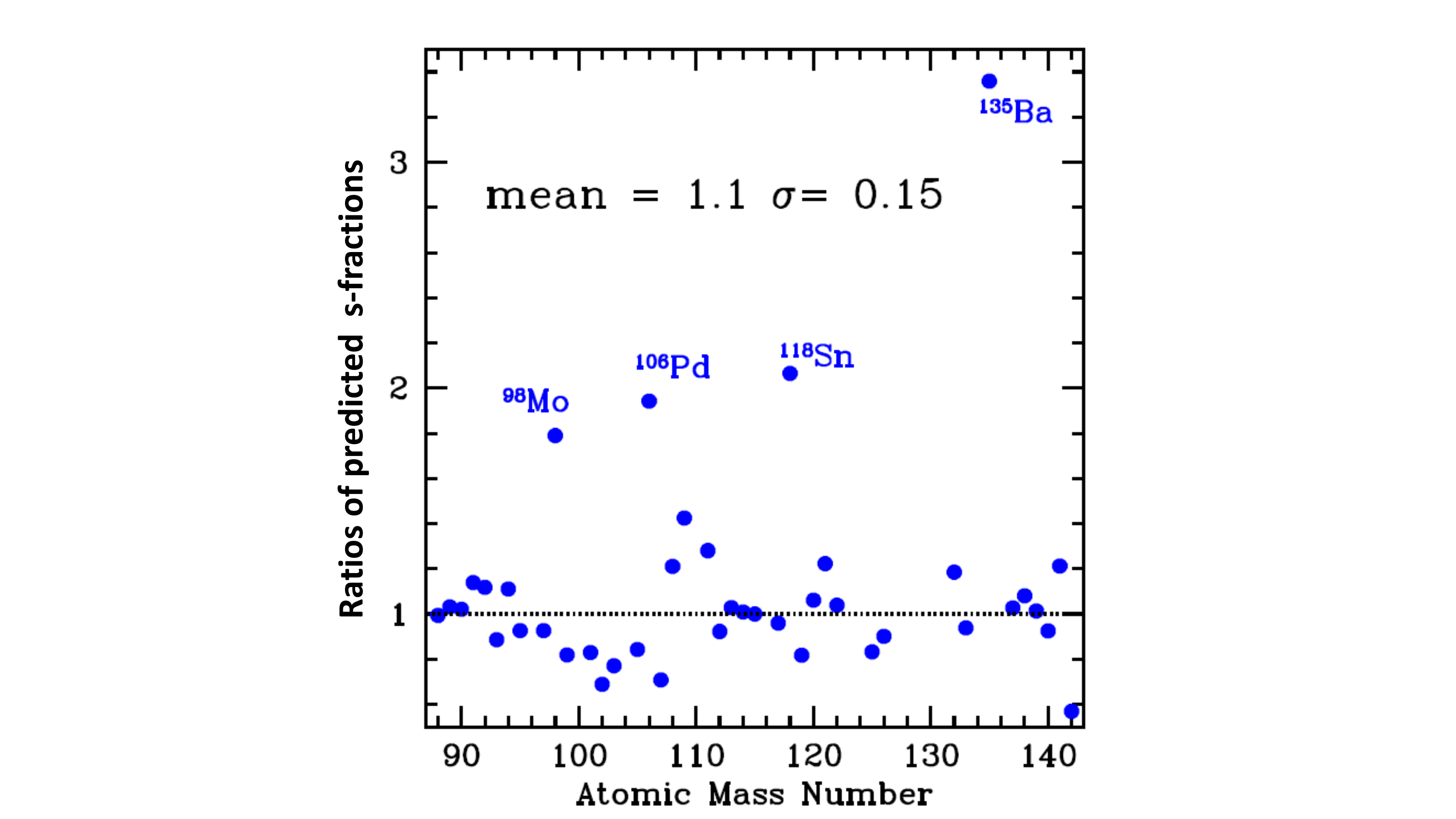}
\caption{Remaining differences between the expectations of solar $s$-process fractions as derived by $s$- and $r$-process models, according to the discussion by \cite{b+22}.}   
\label{fig-7}       
\end{figure}

\begin{figure}[t!]
\includegraphics[width=1.40\textwidth]{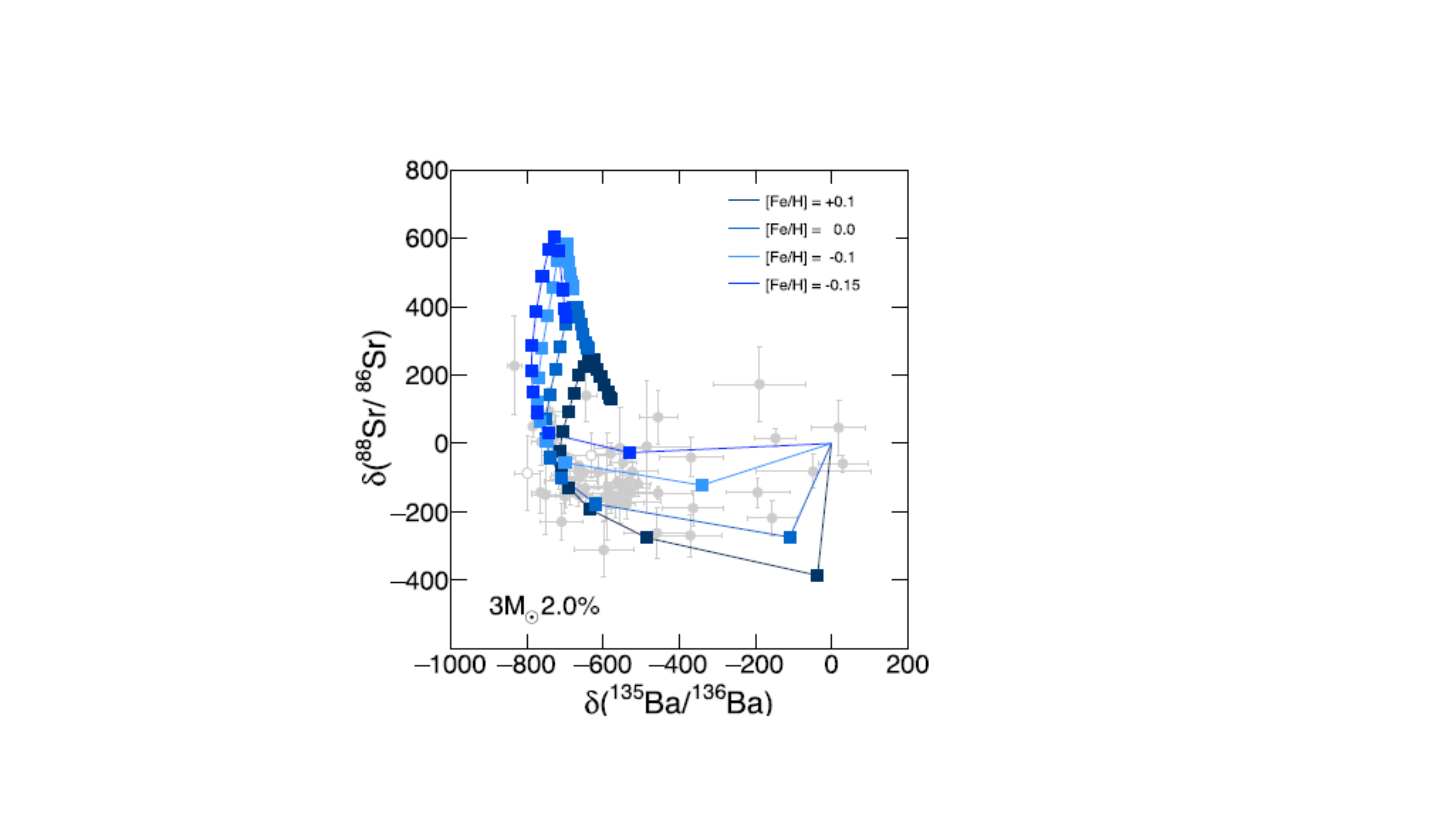}
\caption{The set of isotopic ratios $^{86}$Sr/$^{88}$Sr versus $^{135}$Ba/$^{136}$Ba, expressed in $\delta$-units, as measured in presolar SiC grains (dots with error bars: \cite{liu3}). By comparison, the plot (taken from 
\cite{p+21}) reports predictions from $AGB$ C-stars of originally 3 \msb and various metallicities,
with a further 2\% mixing by flux tubes in the envelope (see the original paper for details).}
\label{fig-8}       
\end{figure}
Displayed in Figure \ref{fig-6} are the logarithms of solar abundance fractions due to $s$-processing, in the mass range 88 $\lesssim $ A $\lesssim$ 142, where the set of nuclear paramters is better constrained by measurements. The plot results from the computations outlined above (see \cite{b+21} for details). The nuclei shielded from fast decays ($s$-only isotopes) are indicated in red. As the figure shows,
models for Galactic $s$-process nucleosynthesis computed with \ctb pockets derived as in Figure \ref{figpock} correctly predict, for them, abundance fractions close to unity. Blue triangles then represent the ensuing $s$-fractions of other nuclei (to be compared with the corresponding expectations from the $r$ process). 

Such a  comparison was performed recently by \cite{b+22}, making use of the site-independent {\it waiting point} approach for the $r$-process, with updated nuclear inputs for masses and decay rates of the relevant
nuclei. Figure \ref{fig-7} shows the remaining discrepancies between the two sets of predictions of the $s$-process fractions of unshielded nuclei. As far as we know, this is the best level of accord published so far. Complementary results, obtained by \cite{p+21} while comparing $AGB$ predictions from 
the above scenario with revised isotopic ratios of heavy nuclei derived from presolar SiC grains
 \cite{liu3}, added further credibility to the global picture (see e.g. Figure \ref{fig-8}).

The models by \cite{tri, b+21, p+21, b+22}) induce remarkable differences with respect to previous coputations. In particular: 
\begin{itemize}
\item{The concentration of $^{19}$F, possibly synthesized from reactions starting at nitrogen, is now 
strongly reduced as compared to exponential-like \ctb pockets (like tha one indicated in Fig. 5, panel i) and also compared to more sophisticated treatments like thos by \cite{c+09}, thus solving the problems previously
affecting this nucleus \cite{v+20, v+21}.}

\item{The new \ctb pockets of the mentioned scenario make in general the nucleosynthesis of heavy nuclei more effective than before. }

\item{The reduced $^{14}$N concentration also induces a lower production of $^{22}$Ne, hence reducing further the efficiency of neutron captures in the convective thermal pulses. Neutrons are now primarily due to the \ctanb reaction, at very low neutron densities ($N_n \simeq 10^7 n/cm^3$). The neutron captures coming out of this model imply in general reaction branchings less effective than before. This fact favors the agreement with presolar grain isotopic ratios (see \cite{p+21}).}

\end{itemize}

All the mentioned facts descend from the type of proton penetration, which is now exponential {\it in density}, not in mass. This originates the quadratic dependence on radius in equation (29), describing a shallow profile, with a local abundance of \ctb much lower, but also much more extended than in previous models (see e.g. \cite{ar+99}). 

The above discussion suggets also how any other hydrodynamical mechanism operating similarly to magnetic buoyancy in the mild $T$ gradient below TDU and introducing forces variable in time, might be suited to induce mixing processes analogous to the ones described here. We hope this fact can encourage other stellar physicists to verify this possibility, thus expanding our understanding of AGB mixing and of the ensuing neutron captures. 
\newpage
{\large\bf {Acknowledgements}}. 

M.B is grateful to the INFN Section of Perugia, to its director, dr. Patrizia Cenci, and to
the authorities of the Institute for continued support after retirement. Useful discussions 
with R. Gallino on the subjects of our longstanding collaboration and friendship and on our
contributions to this volume are gratefully recognized. We are particularly indebted to the 
referees for their very careful analysis and useful suggestions.

\bibliography{BP23-bibliography}

\end{document}